# CaCo$_2$TeO$_6$: A topochemically prepared 3$d^7$ honeycomb Kitaev magnet

Yuya Haraguchi[1,*], Yuto Yoshida[1], Akira Matsuo[2], Koichi Kindo[2], and Hiroko Aruga Katori[1]

[1]*Department of Applied Physics and Chemical Engineering, Tokyo University of Agriculture and Technology, Koganei, Tokyo 184-8588, Japan*
[2]*The Institute for Solid State Physics, The University of Tokyo, Kashiwa, Chiba 277-8581*
**Correspondence author:* chiyuya3@go.tuat.ac.jp

We report the magnetic properties of CaCo$_2$TeO$_6$ as a Kitaev candidate. CaCo$_2$TeO$_6$ was synthesized through a topochemical process, wherein all Na$^+$ ions in Na$_2$Co$_2$TeO$_6$ were replaced with half the amount of Ca$^{2+}$ ions. This substitution brings the CoO$_6$ octahedra closer to an approximate cubic symmetry. CaCo$_2$TeO$_6$ exhibits antiferromagnetic ordering at $T_N \sim 13$ K, which is lower than $\sim 27$ K observed for Na$_2$Co$_2$TeO$_6$. Notably, its magnetic order is suppressed in a relatively low magnetic field of approximately 4 T, indicating that non-Kitaev interactions can be restrained by reducing trigonal distortion. Our findings highlight the potential of CaCo$_2$TeO$_6$ as a viable platform for exploring Kitaev quantum spin liquids and pave the way for a deeper understanding of the fundamental mechanisms in Kitaev physics.

## 1. Introduction

Quantum spin liquids (QSL) have garnered significant attention in condensed matter physics due to their unique and intriguing properties [1-6]. Among the various theoretical frameworks proposed to describe these exotic states of matter, the Kitaev model is a promising avenue for understanding QSL [7-9]. Introduced by Alexei Kitaev in 2006, the Kitaev model is an exactly solvable quantum spin-1/2 model featuring bond-dependent Ising interactions on a honeycomb lattice. This solvability is significant as it offers a comprehensive understanding of the complex behaviors exhibited by QSL.

Furthermore, Kitaev model is particularly fascinating as it gives rise to a QSL ground state [7-9]. This exotic state of matter hosts fractionalized Majorana fermionic excitations, which are intrinsically connected to the $Z_2$ gauge field, and exhibits topological degeneracy that is robust against local perturbations [10-12]. Understanding these features provides valuable insights into the intricate and unconventional properties of QSL.

In the early stages of Kitaev material exploration research, attention was primarily focused on 4$d$/5$d$ transition metal compounds, with a growing interest recently in Co-based quasi-2D honeycomb magnets [13-16]. This shift has been driven by theoretical advancements suggesting that Kitaev interactions between $J_{eff} = 1/2$ pseudospins on Co$^{2+}$ ions can be realized under an octahedral crystal field [13-16]. These interactions are key to understanding the unique magnetic properties of these materials.

Without spin-orbit coupling, Co$^{2+}$ ions exhibit a high-spin configuration (S = 3/2) and possess orbital degrees of freedom ($L_{eff} = 1$) [17,18]. In this scenario, the weak spin-orbit interactions are instrumental in inducing the formation of spin-orbit entangled $J_{eff} = 1/2$ pseudospins, which are critical for generating Kitaev interactions. This nuanced understanding of spin-orbit interactions and their effects on Co$^{2+}$ ions is pivotal in exploring new Kitaev materials.

In the initial studies, Na$_2$Co$_2$TeO$_6$ and Na$_3$Co$_2$SbO$_6$ have been identified as promising candidate materials for realizing Co-based Kitaev magnetism [19-25], exhibiting a zigzag-type magnetic order similar to α-RuCl$_3$ [26], and the low-energy spin dynamics can be described by the Heisenberg-Kitaev-Γ Hamiltonian [27],

$$H = \sum_{\langle i,j \rangle} (J\mathbf{S}_i \cdot \mathbf{S}_i + K S_i^\gamma \cdot S_j^\gamma + \Gamma(S_i^\alpha \cdot S_j^\beta + S_i^\beta \cdot S_j^\alpha)), \quad (1)$$

where $\{i, j\}$ denotes the nearest-neighbor bonds, $\gamma$ takes value $x$, $y$ or $z$ depending on the direction of the nearest-neighbor bond, $\alpha$ and $\beta$ represent the remaining two directions to $\gamma$, $J$ and $K$ are the magnitude of the Heisenberg and Kitaev interactions, and $\Gamma$ characterizes the strength of the off-diagonal exchange interactions. In two-dimensional magnetic systems, the unavoidable trigonal distortion modifies the $J_{eff} = 1/2$ wave function, enhancing non-Kitaev interactions and destabilizing the QSL state. Na$_2$Co$_2$TeO$_6$ displays magnetic order at a relatively high temperature of $T_N = 27.1$ K, indicating that non-Kitaev interactions are sufficiently significant. Furthermore, Na$_2$Co$_2$TeO$_6$ exhibits successive magnetic ordering at even lower temperatures [28], surpassing the scope of the simple Kitaev-Heisenberg-Γ model. To address these challenges, extensive experimental investigations exploring novel approaches grounded in solid-state chemistry are essential for elucidating the relationship between crystal structure and magnetism in Co-based Kitaev materials.

Topochemical reactions offer a versatile approach to manipulating crystal lattice structures by enabling the exchange of ions under mild temperature conditions [29]. This method allows for the formation of novel compounds with finely tuned magnetic states and structural modifications. By applying these techniques to Kitaev iridates, such as α-Li$_2$IrO$_3$ and β-Li$_2$IrO$_3$, novel Kitaev materials have been generated, along with the successful control of magnetic interaction parameters [12, 30-34]. Expanding this method to Co-based Kitaev materials allows the modification of lattice constants across the crystal while preserving the honeycomb network. Concurrently, the crystal field environment surrounding Co$^{2+}$ ions can be changed due to the chemical pressure effect.

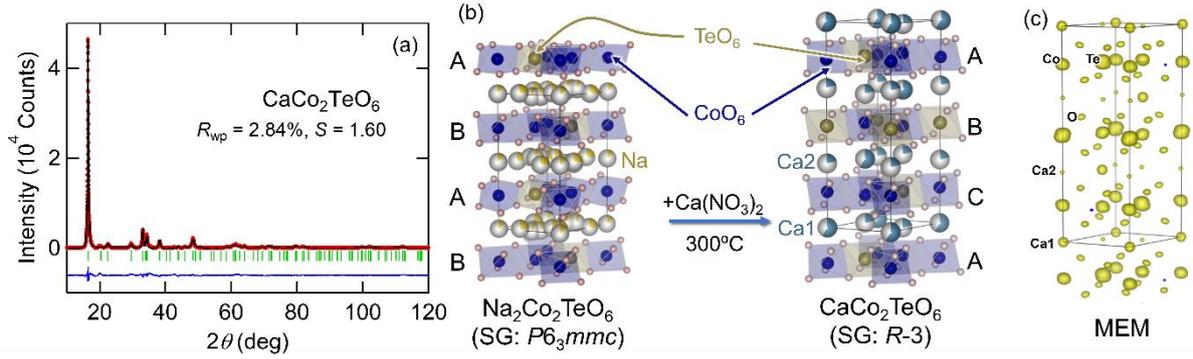

**Figure 1.** (a) Room-temperature powder x-ray diffraction patterns of $CaCo_2TeO_6$. The observed intensities (red), calculated intensities (black), and their differences (blue) are shown. Vertical bars indicate the positions of Bragg reflections. (b) Crystal structure of $CaCo_2TeO_6$, along with a change in stacking patterns through the topochemical process relative to the precursor $Na_2Co_2TeO_6$. The vesta software is used for visualization [36]. (c) Three-dimensional nuclear density distributions of $CaCo_2TeO_6$ calculated by maximum entropy method (MEM) through the z-rietveld software [37].

Moreover, Liu *et al.* have theoretically investigated the detailed superexchange interaction between $CoO_6$ octahedra accounting for the trigonal distortion effect [35], offering microscopic theoretical evidence of two key trends. Firstly, it demonstrates that the Kitaev interaction decreases with increased trigonal distortion. Secondly, it indicates that the off-diagonal interaction escalates with the rise in trigonal distortion. These insights suggest that manipulating the distortion of $CoO_6$ octahedra through local substitution of elements via topochemical reactions could be a promising strategy to realize Kitaev spin liquids.

In this study, $CaCo_2TeO_6$ was successfully synthesized via aliovalent topochemical ion exchange, wherein all $Na^+$ ions in $Na_2Co_2TeO_6$ were entirely replaced by half the amount of $Ca^{2+}$ ions. This substitution brought the $CoO_6$ octahedra closer to cubic symmetry. Moreover, the suppressed magnetic ordering temperature down to $T_N \sim 13$ K and low magnetic field requirement of $\mu_0 H_c \sim 4$ T for a breakdown of magnetic ordering demonstrate the reduction of non-Kitaev terms by limiting trigonal distortion. These findings strongly imply that $CaCo_2TeO_6$ is a promising Kitaev candidate material, demonstrating the potential of topochemical reactions in unlocking avenues for research and material development.

## 2. Experimental Procedure

The precursor $Na_2Co_2TeO_6$ was synthesized through conventional solid-state reactions [38]. All reagents were sourced from Kojundo Chemical Lab. Co., Ltd., with the following purities: calcium nitrate tetrahydrate $(Ca(NO_3)_2 \cdot 4H_2O)$ at 99.9%, tellurium dioxide ($TeO_2$) at 99.9%, cobalt(II) oxide (CoO) at 99.7%, and sodium carbonate ($Na_2CO_3$) at 99%. For calcium nitrate, the tetrahydrate form was dehydrated at 400°C prior to use to ensure its suitability for the following experimental procedures. Stoichiometric amounts of $Na_2CO_3$, $TeO_2$, and CoO were mixed, and the mixture was calcined at 900°C for 96 h in air. This precursor was ground well with a large excess of $Ca(NO_3)_2$ (m.p. 561 °C) in an Ar-filled glove box, sealed in an evacuated Pyrex tube, and reacted at 300°C for two days. The ion-exchange reaction is expressed as,

$$Na_2Co_2TeO_6 + Ca(NO_3)_2 \rightarrow CaCo_2TeO_6 + 2NaNO_3. \quad (2)$$

After the reaction, the pellets were partially melted due to the formation of eutectic salts (minimum m.p. 225 °C) of the byproducts $NaNO_3$ and unreacted $Ca(NO_3)_2$. The observed melting indicates that the reaction in Eq.(2) was promoted, and the byproduct $NaNO_3$ precipitated. The eutectic salts were removed by repeatedly washing the sample with distilled water. The obtained polycrystalline samples were characterized by powder x-ray diffraction (XRD) experiments in a diffractometer with Cu-Kα radiation. The cell parameters and crystal structure were refined using the Rietveld method with the Z-RIETVELD software [37]. The electron density distribution was calculated using the maximum-entropy method (MEM) through the Z-RIETVELD software [37].

**Table 1.** Crystallographic parameters for $CaCo_2TeO_6$ ($R\bar{3}$) determined from powder x-ray diffraction experiments. The obtained lattice parameters are $a = 5.22585(8)$ Å and $c = 16.2312(1)$ Å. $B$ is the atomic displacement parameter.

| Atom | Site | Occupancy | x | y | z | B (Å) |
|---|---|---|---|---|---|---|
| Co | 6c | 1 | 0 | 0 | 0.17725(9) | 0.74(5) |
| Te | 3b | 1 | 0 | 0 | 1/2 | 0.12(3) |
| Ca1 | 3a | 0.574(4) | 0 | 0 | 0 | 0.24(5) |
| Ca2 | 6c | 0.213(2) | 0 | 0 | 0.31706(3) | 0.24(5) |
| O | 18f | 1 | 0 | 0.3080(5) | 0.09453(8) | 0.77(9) |

The temperature dependence of the magnetization was measured under several magnetic fields using the magnetic property measurement system (MPMS; Quantum Design) at the Institute for Solid State Physics (ISSP), the University of Tokyo. The temperature dependence of the heat capacity was measured using the conventional relaxation method in a physical property measurement system (PPMS; Quantum Design) at ISSP, the University of Tokyo. Hot-pressed powdered $CaCo_2TeO_6$ samples were prepared for heat capacity measurements by superheating at 300°C for 12 hours under approximately 100 kg of pressure using a vise-heat press, to optimize thermal contact during the measurements. Magnetization curves up to approximately 25 T were measured by the induction method in a multilayer pulsed magnet at the International Mega Gauss Science Laboratory at ISSP, the University of Tokyo.

## 3. Results

The room-temperature powder X-ray diffraction pattern of $CaCo_2TeO_6$ is presented in Fig. 1(a). The peaks can be consistently assigned to reflections based on the space group of $R\bar{3}$ with trigonal lattice constants of $a$ = 5.225 85(8) Å and $c$ = 16.2312(1) Å. This space group is distinct from that of the precursor $Na_2Co_2TeO_6$, indicating a change in the stacking pattern of the Co honeycomb layers from AB stacking to ABC stacking as a result of the ion exchange process, as shown in Fig. 1(b). The chemical composition was analyzed through energy-dispersive X-ray spectrometry, which revealed that Co/Ca = 2.00(8), confirming the successful completion of the ion exchange reaction from $Na_2Co_2TeO_6$ to $CaCo_2TeO_6$. The $R\bar{3}$ structure model (similar to the ilmenite-type structure) perfectly reproduce the XRD profile. The crystal structure of $CaCo_2TeO_6$ was refined using the Rietveld method, as detailed in the Experimental Methods section. The refined crystallographic parameters are summarized in Table I, and the visualized structure is shown on the right side of Fig. 1(b). The bond valence sum calculation performed on the refined structural parameters of the Co ions yielded a value of +2.084. This value is consistent with the expected valence of +2, thereby providing further support for the validity of the structure [39]. Moreover, the electron density distribution for the unit cell of $CaCo_2TeO_6$, obtained from implementing MEM calculations and depicted in Fig. 1(c), was employed to assess the veracity of the crystal structure of $CaCo_2TeO_6$. The remarkable congruence between the electron density magnitude and spatial arrangement with the structural

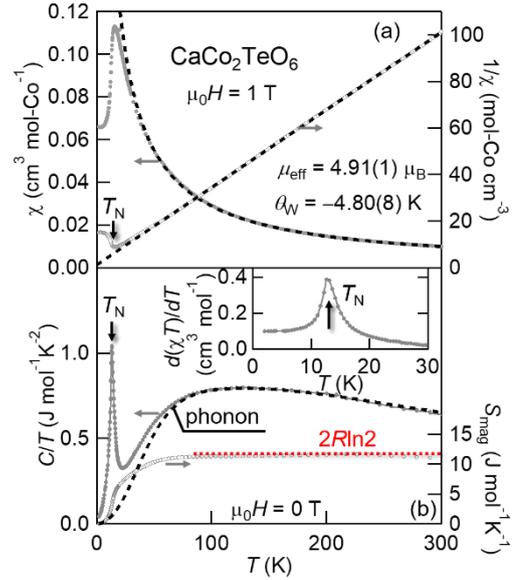

**Figure 2.** (a) Temperature dependence of magnetic susceptibility χ and its inverse 1/χ in powder sample of $CaCo_2TeO_6$. Broken lines indicate the Curie-Weiss fits. Magnetic transition at $T_N$ is highlighted. (b) Temperature dependence of the heat capacity divided by temperature $C/T$ and the estimated magnetic entropy $S_{mag}$ in $CaCo_2TeO_6$ under zero magnetic-fields. Magnetic transition at $T_N$ is highlighted. The black broken line indicated the phonon contribution estimated by fitting the data above 100 K, as described in the text. The horizontal dashed lines indicate the value of $S_{mag}$ = $2R\ln 2$, which is the total magnetic entropy derived from $J_{eff}$ = 1/2. The inset shows the $d(\chi T)/dT$ data referred to as so-called Fisher heat capacity [45].

analysis results unequivocally validates the $R\bar{3}$ structure.

Figure 2(a) shows the temperature dependences of magnetic susceptibility χ and its inverse 1/χ for the polycrystalline $CaCo_2TeO_6$ measured at $\mu_0H$ = 1 T. A Curie-Weiss fitting of the 1/χ data at 200–300 K yields an effective magnetic moment $\mu_{eff}$ = 4.91(1) $\mu_B$ and Weiss temperature $\theta_W$ = −4.80(8) K. The observed enhancement of $\mu_{eff}$ relative to the expected value of 3.87 $\mu_B$ for a spin-only scenario ($S$ = 3/2) indicates coupling with an unquenched orbital angular momentum of $L_{eff}$ = 1. This observation aligns with the established $\mu_{eff}$ values observed for well-investigated $Co^{2+}$-magnets [40], including honeycomb [19-21,38,41-42], kagome [43], and triangular systems [44]. The magnetic susceptibility peaks at around 16 K, indicating a sharp drop in susceptibility just below this temperature, suggesting an antiferromagnetic transition. The Néel temperature is determined to be 13 K, as evidenced by the peak-top in the $d(\chi T)/dT$ data (see the inset of Figure 2(b)), referred to as so-called Fisher heat capacity, which is the relationship between magnetic heat capacity and magnetic susceptibility $C_p(T) \propto \partial(\chi T)/\partial T$ [45]. This conclusion is evidenced by observing a λ-type peak in the heat capacity divided by temperature $C/T$ at $T_N$ as shown in Figure 2(b). The $T_N$ value, which exceeds the absolute $|\theta_W|$ value, suggests the coexistence of comparable ferromagnetic and antiferromagnetic interactions. Based on the observation, it is reasonable to conclude that the ferromagnetic and antiferromagnetic interactions are in competition with each other.

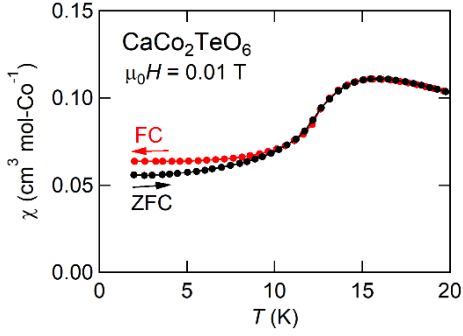

**Figure 3** Temperature dependence of magnetic susceptibility of $CaCo_2TeO_6$ measured during heating after zero-field cooling (ZFC), followed by cooling (FC) under an applied magnetic field of 0.01 T.

To accurately quantify the entropy release associated with the long-range magnetic order in $CaCo_2TeO_6$, it is crucial to differentiate between the magnetic heat capacity $C_{mag}$ and the phonon heat capacity $C_{phonon}$. This is achieved by subtracting $C_{phonon}$ from the total heat capacity. Efforts to synthesize $CaZn_2TeO_6$, as a potential non-magnetic reference compound for $CaCo_2TeO_6$, using a topochemical reaction similar to that used for $Na_2Zn_2TeO_6$ were unsuccessful. It is well known that a lattice heat capacity $C_{phonon}$ consists of contributions from three acoustic phonon branches and $3n-3$ optical phonon branches, where $n$ is the number of atoms per formula unit [46]; $n$ equals 10 for $CaCo_2TeO_6$. The acoustic and optical contributions are described by the Debye- and Einstein-type heat capacities $C_D$ and $C_E$, respectively. Provided that $C_{phonon}$ is the sum of $C_D$ and $C_E$, the $C/T$ data above 100 K, where the magnetic heat capacity may be negligible, are reproduced by the equation,

$$C_{lattice} = C_D + C_E$$
$$= 9R(T/\theta_D)^3 \int_0^{\theta_D/T} \frac{x^4 \exp(x)}{[\exp(x)-1]^2} dx$$
$$+ R\sum_{i=1}^{3} n_i \frac{(\theta_{Ei}/T)^2 \exp(\frac{\theta_{Ei}}{T}-1)}{\exp(\frac{\theta_{Ei}}{T}-1)} \quad (3)$$

, where $R$ is the gas constant, $\theta_D$ is the Debye temperature, $\theta_{Ei}$ is the Einstein temperatures. The best fits are shown by the dashed lines with $\theta_D = 277$ K, $\theta_{E1} = 325$ K, $\theta_{E2} = 758$ K, $\theta_{E3} = 1890$ K, $n_1 = 6$, $n_2 = 15$, and $n_3 = 6$ for $CaCo_2TeO_6$. The

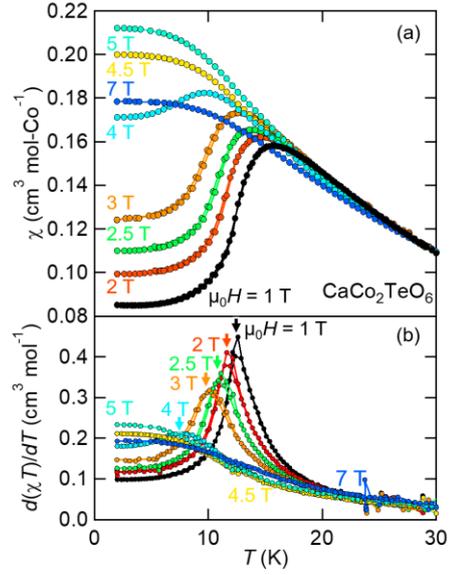

**Figure 4** (a) The temperature dependences of χ measured for several magnetic fields for $CaCo_2TeO_6$. (b) $d(\chi T)/dT$ data. Arrows indicate the positions of magnetic anomalies.

magnetic contribution $C_{mag}/T$ was obtained by subtracting this lattice contribution determined by $C/T$ of the estimated $C_{phonon}/T$. Then, the magnetic entropy $S_{mag}$ was calculated by integrating $C_{mag}/T$ with respect to $T$. As shown in Fig. 2(b), the obtained $S_{mag}$ approaches $2R\ln 2 = 11.52$ J mol$^{-1}$K$^{-1}$ expected for a doublet, demonstrating that the $J_{eff} = 1/2$ doublet is realized. The $S_{mag}$-value reaches approximately 2.92 J mol$^{-1}$K$^{-1}$ at $T_N$, accounting for about 25% of the total magnetic entropy. This suggests that a significant portion of the magnetic entropy is released through the development of short-range magnetic correlations above $T_N$.

At the lowest applied field of 0.01 T, the χ-data displays a slight thermal hysteresis between the zero-field-cooled and field-cooled measurements below the Néel temperature ($T_N = 13$ K), suggestive of a weak spin-glass-like contribution, as shown in Fig. 3. Nevertheless, this discrepancy is very small when considering the total magnetization at 2 K and is markedly smaller than the weak-ferromagnetic thermal hysteresis observed in $Na_2Co_2TeO_6$ [19, 22]. Furthermore, at higher magnetic fields above 1 T, the glassy component becomes nearly undetectable. Therefore, the glass-like behavior is attributed to the freezing of a minority spin population.

Figures 4 expands the low temperature region of χ measured under magnetic fields from 1 to 7 T. At ~ 4 T, the $M$ curve exhibits a metamagnetic-like increase. The temperature at which a sharp drop in the magnetic susceptibility χ data occurs due to magnetic ordering shifts towards lower temperatures with increasing applied magnetic field and disappears at 4.5 T. This behavior is substantiated by the peaks in the $d(\chi T)/dT$ data, which exhibit a broad peak near 7.8 K at 4 T, but the peak vanishes

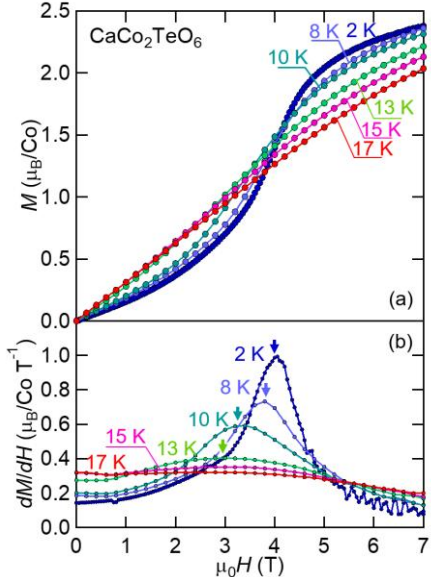
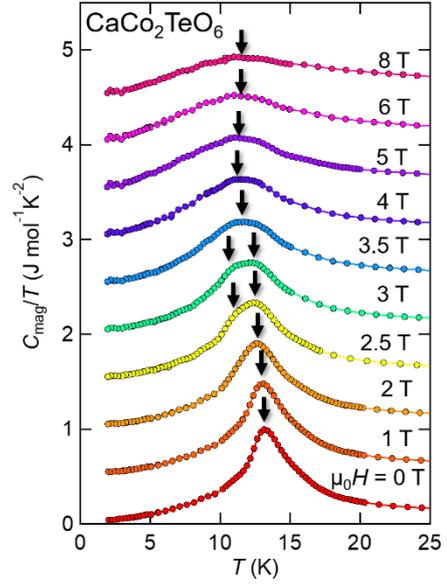

**Figure 5** (a) Isothermal magnetization curves measured at several temperatures for $CaCo_2TeO_6$. (b) $dM/dH$ data. Arrows indicate the positions of magnetic anomalies.

**Figure 6.** Magnetic heat capacity of $CaCo_2TeO_6$ as a function of temperature measured under various magnetic fields. The data are offset by 0.5 J mol$^{-1}$K$^{-2}$ for each successive magnetic field for improved visibility. Arrows indicate the positions of magnetic anomalies.

at 4.5 T. This field dependence indicates that the magnetic ordering is suppressed by the magnetic field.

Figure 5 shows the isothermal magnetization process $M$ at various temperatures. At 2 K, the $M$ data shows a rapid increase in magnetization around 4 K, evidencing a field-induced phase transition at 4 T. The observed increase in magnetization is due to the magnetic-field-induced suppression of magnetic order, a well-documented phenomenon in other Kitaev materials such as α-RuCl$_3$ [26, 47] and $Na_2Co_2TeO_6$ [19, 22]. As temperature increases, the magnetization's rate of change along the field-induced phase transition curve becomes more gradual. Concurrently, the critical field for the phase transition is reduced, as evidenced by the peak of the $dM/dH$ curve shifting towards lower magnetic fields with rising temperature. The magnetic transition temperature, determined from temperature sweep susceptibility measurements at low magnetic fields, is approximately $T_N \sim 13$ K. Field-induced magnetic order suppression should not occur in isothermal magnetization processes above $T_N$ due to the paramagnetism. Contrary to this assumption, the $dM/dH$ curve still exhibits a broad peak ~ 3 T even above 13 K. Notably, the $dM/dH$ peak position remains relatively constant at ~ 3 T as temperature increases, although the peak's intensity gradually diminishes and is eventually suppressed. This suggests the occurrence of some form of crossover phenomenon, which would be attributed by the melting of entangled short-range ordering. The origin of this crossover-like behavior will be discussed later. Unfortunately, obtaining accurate data on the magnetic field dependence of heat capacity has proven challenging. When employing the thermal relaxation method, cold pressing the samples to enhance the contact area with the sample stage causes significant preferential orientation of the samples.

This makes it difficult to acquire reliable heat capacity data for the powder average. Conversely, when the powder is embedded in grease without cold-pressing, the thermal contact is significantly reduced, leading to unreliable data.

Figure 6 shows the low temperature region of magnetic heat capacity $C_{mag}/T$, obtained by subtracting the phonon contribution using the aforementioned method, measured under various magnetic fields. As the applied magnetic field increases, the transition temperature shifts to lower values, with the anomaly disappearing entirely above 3.5 T. Above 3.5 T, the $C/T$ data exhibits a broad peak, with the transition temperature increasing with the magnetic field, characteristic of ferromagnetic behavior. This transformation arises from the destabilization of the antiferromagnetic phase and the energetic stabilization of the ferromagnetic phase, where spins progressively align at higher temperatures under the influence of the applied magnetic field. This behavior mirrors that observed in related cobaltates, such as precursor $Na_2Co_2TeO_6$ [21], $Li_3Co_2SbO_6$ [48], and $BaCo_2(P_{0.85}V_{0.15})_2O_8$ [49]. Moreover, the heat capacity data at 2.5 T and 3 T exhibit a merged peak structure, as indicated by double arrows. This behavior suggests that the two set of phase boundaries are closely spaced or possibly overlapping at these magnetic fields. Consequently, this leads to a more continuous or broadened transition between these phases.

To further elucidate the magnetic states under high magnetic fields, magnetization was measured up to 25 T utilizing pulsed magnetic fields, as shown in Figure 7.

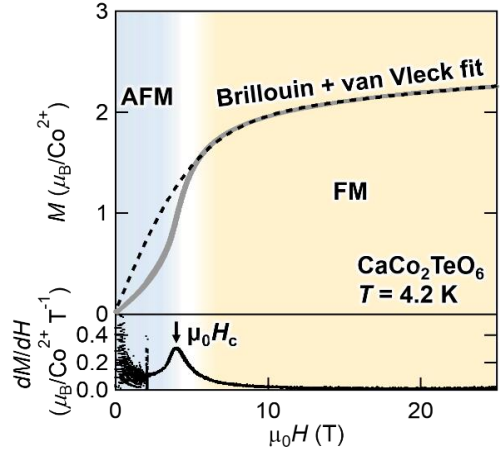

**Figure 7.** The isothermal magnetization curve $M$ and its derivative $dM/dH$ measured at $T = 4.2$ K under pulsed magnetic fields up to 25 T for $CaCo_2TeO_6$. The black dashed line represents the best fit by Eq. 4 described in the text.

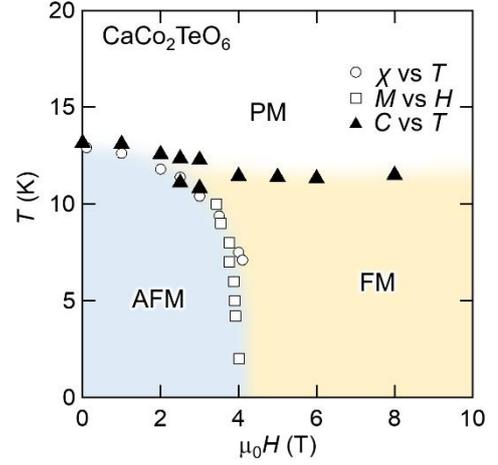

**Figure 8.** Temperature vs magnetic-field phase diagram for $CaCo_2TeO_6$ constructed by the measurement data of magnetization and heat capacity. The following abbreviations are used: PM refers to the paramagnetic phase, AFM to the antiferromagnetic phase, and FM to the ferromagnetic phase.

Above 5.5 T, the magnetization curve is convex upward, showing a gradual increase, whereas above 15 T, it transitions to an almost linear increment with respect to the magnetic field. The initial increase can largely be attributed to the Brillouin function's contribution, which arises from the suppression of magnetic order and a resultant paramagnetic state. In contrast, the linear behavior at higher fields is predominantly governed by van Vleck paramagnetism associated with $Co^{2+}$ spins. By fitting the magnetization curves linearly above 20 T and extrapolating to zero magnetic field, the van Vleck contribution was isolated, providing an approximate saturation magnetization of about $2\mu_B$. The obtained saturation magnetization $M_{sat} = gJ$ of $\sim 2\mu_B$ is characteristic of $Co^{2+}$ magnets, a consequence of the $J_{eff} = 1/2$ pseudospin state with the powder-averaged g-factor of $g_{powder} \sim 4$. Here, in order to separate the contribution of $J_{eff} = 1/2$ pseudospin and the van Vleck term, we analyze the $M$ data with a modified Brillouin function,

$$M = g_{powder} J \mu_B B_S \left( \frac{gJ\mu_B H}{k_B T} \right) + \chi_{vv} H, \quad (4)$$

where $g_{powder}$ is the powder-averaged g-factor, $\mu_B$ is the Bohr magneton, $J (= 1/2)$ is the total angular momentam, $k_B$ is the Boltzmann constant. For the second term, $\chi_{vv}$ is the van Vleck term. The best fit is shown by the dashed line in Figure 7, with the fitting parameters of $g_{powder} = 4.349(1)$ and $\chi_{vv} = 0.0156$ $(\mu_B/T)/Co^{2+} = 0.00872$ cm$^3$/mol-$Co^{2+}$. The estimated $g_{powder}$-value satisfies the criteria for $J_{eff} = 1/2$ ground state [18,43,50], and the $\chi_{vv}$-value is typical for $Co^{2+}$ in an octahedral coordination, a characteristic commonly observed in magnetic systems involving $Co^{2+}$ [51]. Therefore, the good reproduction of the magnetization process above 5.5 T suggests that the $J_{eff} = 1/2$ pseudospins reach saturation and become fully polarized within the forced ferromagnetic regime at fields above the field-induced phase transition field of 4 T.

## 4. Discussion

As previously mentioned, $CaCo_2TeO_6$ is extensively characterized as a honeycomb lattice magnet comprised of $J_{eff} = 1/2$ pseudospins.

First, we summarize a magnetic phase diagram of $CaCo_2TeO_6$, constructed based on magnetic susceptibility, isothermal magnetization, and specific heat data, as shown in Fig. 8. The magnetization data delineates the phase boundary between the antiferromagnetic (AFM) and ferromagnetic (FM) states, with evidence of a metamagnetic transition. As shown in Fig. 7, the High-field magnetization measurement demonstrates that the $Co^{2+}$ ions with an effective spin of $J_{eff} = 1/2$ are fully polarized, eliminating the possibility of ferrimagnetism or canted antiferromagnetism in FM phase.

Next, we discuss the suppression of the magnetic ordering temperature through the topochemical conversion process from $Na_2Co_2TeO_6$ to $CaCo_2TeO_6$ in terms of the local crystal structure. As the extent of trigonal distortion energy increases, the magnitude of the magnetic anisotropy is correspondingly amplified. As shown in Fig. 9, the topochemical transformation from $Na_2Co_2TeO_6$ to $CaCo_2TeO_6$ results in an extension of the upper and lower oxygen planes along the c-axis in the $CoO_6$ octahedron, increasing from 2.135 Å to 2.341 Å, suggesting a corresponding decrease in trigonal distortion. The degree of trigonal distortion in $CoO_6$ octahedra can be quantified using the bond angle variance [52],

$$\sigma^2 = \sum_{i=1}^{m} \frac{(\varphi_i - \varphi_0)^2}{m-1}, \quad (5)$$

where $m$ is the number of anion-cation-anion bond angles; $m = 12$ for octahedra, $\varphi_i$ is the $i$th bond angle of the distorted coordination-polyhedra, and $\varphi_0$ is the ideal bond angle of a polyhedra with $O_h$ symmetry; $\varphi_0 = 90$ deg for octahedra. The $\sigma^2$ value for $CaCo_2TeO_6$ is 36.0220 deg$^2$, which is notably smaller than those of $Na_2Co_2TeO_6$ (60.4403 and 62.4478 deg$^2$ for two Co-sites, calculated from Ref. [53]). Another metric for distortion in $CoO_6$ octahedra is the quadratic elongation [52] defined as,

$$\langle \lambda \rangle = \sum_{i=1}^{6} \frac{l_i/l_0}{n}, \quad (6)$$

where $n$ is the coordination number of anions around the central cation, $l_i$ is the bond length between the central cation and the $i$-th coordinating anions, and $l_0$ is the bond length in a polyhedron with $O_h$ symmetry, with a volume equal to that of the distorted polyhedron. The $\langle \lambda \rangle$-value of $CaCo_2TeO_6$ is 1.0104, which is smaller than that of $Na_2Co_2TeO_6$ (1.0180 for both Co-sites, calculated from Ref. [52]). These structural distortion metrics are attributed to the compression of the $CoO_6$ octahedron within the ab-plane, driven by a reduction in the $a$-axis lattice parameter from 5.289 Å to 5.225 Å, which in turn induces an expansion of the $CoO_6$ octahedron along the $c$-axis. This contraction is induced by the substitution of interlayer ions. As a compensatory effect, the $CoO_6$ octahedron undergoes elongation along the $c$-axis, ensuring the maintenance of the overall valence sum of the Co-O bonds as shown in Fig. 9. Thus, the transformation from $Na_2Co_2TeO_6$ to $CaCo_2TeO_6$ through topochemical ion exchange leads to the suppression of trigonal distortion, resulting in a lower magnetic transition temperature and a smaller magnetic field required to suppress the magnetic transition. These changes indicate a relative increase in the Kitaev interaction compared to the Heisenberg and Γ interactions in Eq. (1), aligning with the theoretical calculations predicted by Liu et al. [34], and providing a clear link between structural modifications and magnetic behavior.

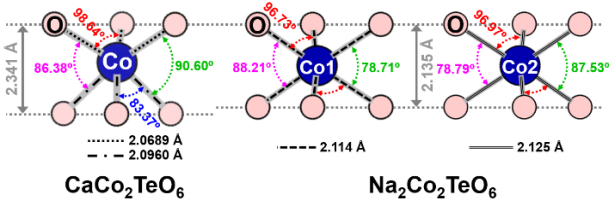

**Figure 9.** Comparison of local structural parameters of $CoO_6$ octahedra between $CaCo_2TeO_6$ and $Na_2Co_2TeO_6$.

These results indicate the potential for tuning Hund's coupling and trigonal distortion to realize the Kitaev QSL by optimizing the interlayer-ions through topochemical reactions. For instance, inserting smaller ions such as Cd or Mg, instead of Ca, may be advantageous. However, the reduction in interlayer distance might increase interlayer interactions, potentially destabilizing the QSL. Alternatively, introducing asymmetric ions, such as organic molecule cations, could minimize interlayer interactions while simultaneously reducing trigonal distortion through compression in the $ab$-plane direction. Furthermore, selecting a substrate capable of compressing the $ab$-plane direction to create epitaxial thin films could achieve the optimal crystal structure for realizing the QSL.

## 5. Summary

We successfully synthesized $CaCo_2TeO_6$ by employing a topochemical reaction between $Na_2Co_2TeO_6$ and calcium nitrate. This process reduced the trigonal crystal field environment surrounding the $Co^{2+}$ ions. Because of this modification, there was a notable reduction in the magnetic ordering temperature, accompanied by the elimination of successive phase transition behavior. Furthermore, the magnetic order in the synthesized $CaCo_2TeO_6$ was suppressed under a relatively low magnetic field. This observation suggests that non-Kitaev interactions can be effectively restrained by reducing trigonal distortion. The magnetic properties of $CaCo_2TeO_6$ exhibit characteristics of $J_{eff} = 1/2$ honeycomb lattice magnets to explore a possible $Co^{2+}$-based Kitaev magnetism.

**Acknowledgments**

This work was supported by JST PRESTO Grant Number JPMJPR23Q8 (Creation of Future Materials by Expanding Materials Exploration Space) and JSPS KAKENHI Grant Numbers. JP23H04616 (Transformative Research Areas (A) "Supra-ceramics"), JP22K14002 (Young Scientific Research), and JP21K03441 (Scientific Research (C)). Part of this work was carried out by joint research in the Institute for Solid State Physics, the University of Tokyo (Project Numbers 202012-HMBXX-0021, 202112-HMBXX-0023, 202106-MCBXG-0065 and 202205-MCBXG-0063).